\documentclass{elsart41}
\usepackage{graphics}
\usepackage{graphicx}
\usepackage{amssymb}
\newcommand{\beq}{\begin{equation}}
\newcommand{\eeq}{\end{equation}}
\begin{document}

\begin{frontmatter}



\title{Quasiparticles near quantum phase transition
in heavy fermion metals}
%

\author[*]{V.R. Shaginyan},
\ead{vrshag@thd.pnpi.spb.ru}

\address{Petersburg Nuclear Physics Institute, RAS, Gatchina 188300, Russia}

\corauth[V.R. Shaginyan]{Corresponding author}

\begin{abstract}

We have shown that the Landau paradigm based upon both the
quasiparticle concept and the notion of the order parameter
can be used to explain the anomalous behavior of
heavy fermion metals. Exploiting this
paradigm and the fermion condensation quantum phase transition (FCQPT)
we show that this anomalous behavior is universal
and can be used to capture the essential aspects of recent
experiments on the heavy-fermion metals at low temperatures.
Behind FCQPT a tunneling conductivity between a heavy fermion metal and a simple metallic
point can be noticeably dissymmetrical with respect to the change of
voltage bias. We show that at $T=0$ and beyond FCQPT the Hall coefficient undergoes a jump
upon magnetic-field tuning HF metals.

\end{abstract}

\end{frontmatter}


Experiments on heavy-fermion (HF) metals explore
mainly their thermodynamic properties which proved to be quite
different from that of ordinary metals described by the Landau
Fermi liquid (LFL) theory. In the LFL theory, considered as the
main instrument when investigating quantum many electron physics,
the effective mass $M^*$ of quasiparticle excitations controlling
the density of states determines the thermodynamic properties of
electronic systems. It is possible to explain many of the observed
properties of the HF metals on the basis of FCQPT which is in harmony with
the Landau paradigm resting upon both
the quasiparticle concept and the notion of the order parameter.
In contrast to the conventional Landau quasiparticles, these are
characterized by the effective mass
which  strongly depends on temperature $T$, applied magnetic field
$B$ and the number density $x$ of the heavy electron liquid of HF
metal.

To study the universal behavior of the HF metals at low temperatures
we use a heavy electron liquid model in order to get rid of the specific peculiarities of HF metals. Since we consider processes related to the power-low divergency of the effective mass, this divergency is determined by small momenta transferred as compared to momenta of the order of the reciprocal lattice, and the contribution coming from the lattice can be ignored.
We start with the well-known Landau
equation determining $M^*(T,B)$
\beq \frac{1}{M^*(T,B)}=
\frac{1}{M}
+\int {\bf Y}
F({\bf p_F},{\bf p}_1) \frac{\partial
n({\bf p}_1)}{\partial {p}_1} \frac{d{\bf
p}_1}{(2\pi)^3}. \eeq For brevity, we omit  spin variables.
Here ${\bf Y}=({\bf p}_F{\bf p_1})/{p_F^3}$,\,
$F({\bf p_F},{\bf p}_1)$ is the Landau amplitude depending on
the momenta $p$, $p_F$ is the Fermi momentum, $M$ is the bare mass
of an electron, and $n({\bf p}_1,T,B)$ is the quasiparticle
distribution function. Applying Eq. (1) at $T=0$ and $B=0$, we obtain the
standard result
$ M^*(x)=M/(1-N_0F^1(p_F,p_F)/3).$
Here $N_0$ is the density of states of the free Fermi gas and
$F^1(p_F,p_F)$ is the $p$-wave component of the Landau
interaction. Since in the LFL theory $x=p_F^3/3\pi^2$, the Landau
amplitude can be written as $F^1(p_F,p_F)=F^1(x)$. Assume that at
some critical point $x_{FC}$ the denominator
$(1-N_0F^1(x)/3)$ tends to zero, and one obtains
that $M^*(x)$ behaves as
$M^*(x)/M\propto1/r$.
Here $r=(x-x_{FC})$ is the ``distance'' from the QCP of FCQPT
taking place at $x_{FC}$.

At $r\to 0$ and $T<T^*(B-B_{c0})\propto (B-B_{c0})$, the system has the LFL behavior
and a qualitative analysis of Eq. (1) shows that
$M^*(B)\propto (B-B_{c0})^{-2/3}.$ Here,
$B_{c0}$ is the critical magnetic field which drives both a HF
metal to its magnetic field tuned QCP and the corresponding N\'eel
temperature $T_N\to 0$ \cite{shag4}. At elevated temperatures the system demonstrates two types of the non-Fermi liquid behavior (NFL): at $T\sim T^*(B)$
$M^*(T)\propto T^{-2/3},$ and $T> T^*(B)$
$M^*(T)\propto 1/\sqrt{T}.$ These mentioned regimes can
be observed in measurements of the resistivity,
$\rho(T)=\rho_0+\Delta\rho$. Here, $\rho_0$
is the residual resistivity, $\Delta\rho=A(B)T^2$, and
$A(B)\propto (M^*)^2$. The first LFL
regime is represented by
$\Delta\rho_1\propto T^2/(B-B_{c0})^{4/3}\propto T^2$; the second NFL
one is characterized by
$\Delta\rho_2\propto T^2/(T^{2/3})^2\propto T^{2/3}$; and the third
NFL one is represented by
$\Delta\rho_3\propto T^2/(\sqrt{T})^2\propto T$. All these regimes were
observed in recent measurements on the HF metals, see e.g. \cite{pag}.
Considering the ratio
$\Delta\rho_2/\Delta\rho_1\propto ((B-B_{c0})/T)^{4/3}$, we conclude
that the ratio is a function of
only the variable $(B-B_{c0})/T$ representing the scaling
behavior. This result is in excellent agreement with experimental
facts \cite{pag}.

Beyond the critical point $x_{FC}$ the distance $r$ becomes negative making
the effective mass negative. To escape the
possibility of being in unstable and meaningless states with the
negative effective mass, the system is to undergo FCQPT
quantum phase transition at the critical point $x=x_{FC}$ \cite{shag4}.
At $x<x_{FC}$ the
quasiparticle distribution is determined by the standard equation
to search the minimum of a functional
${\delta E[n({\bf p})]}/{\delta n({\bf
p},T=0)}=\varepsilon({\bf p})=\mu; \,p_i\leq p\leq p_f.$
This equation determines the quasiparticle distribution function
$n_0({\bf p})$ which delivers the minimum  value to the ground
state energy $E$. $n_0({\bf p})$ does not coincide with the step function in the
region $(p_f-p_i)$, so that $0<n_0({\bf p})<1$,while $p_i<p_F<p_f$.
This special behavior of $n_0({\bf p})$
determines the behavior of system at $T<T_f$, with $T_f$ being the temperature at which he influence FCQPT vanishes. At $T=0$,
the relevant order parameter is the superconducting-like, $\kappa({\bf
p})=\sqrt{n_0({\bf p})(1-n_0({\bf p}))}$, with the entropy $S=0$ \cite{shag4}. At $0<T<T_f$,
the entropy can be approximated as
$S(T) \simeq S_0+a\sqrt{T/T_f}$, $a$ is a constant. This temperature independent term $S_0$ determines the specific NFL behavior of the system. For example, the thermal expansion coefficient $\alpha(T)$ determined by the contribution coming from $S_0$  becomes constant at $T\to0$ while the specific heat $C\sim a\sqrt{T/T_f}$. As a result, the
Gr\"uneisen ratio $\Gamma$ diverges $\Gamma=\alpha/C\propto 1/\sqrt{T}$.  Then, at $T<T_f$, a tunneling conductivity between a heavy fermion metal and a
simple metallic point can be noticeably dissymmetrical
with respect to the change of voltage bias \cite{tun}. At $T<T^*(B-B_{c0})\propto \sqrt{B-B_{c0}}$, the application of magnetic field $B$
restores the LFL behavior, the effective mass becomes
$M^*(B)\propto 1/\sqrt{B-B_{c0}}$, and the coefficient $A\propto1/(B-B_{c0})$ \cite{shag4}.
At $T=0$, the application of the critical magnetic field $B_{c0}$ suppressing the antiferromagnetic state (AF)(with the Fermi momentum $p_{AF}\simeq p_F$) restores the LFL with the Fermi momentum $p_f>p_F$. Both AF and LFL have the same ground state energy being degenerated at $B=B_{c0}$. Thus, at $T=0$ and $B=B_{c0}$, the infinitesimal change in the magnetic field $B$ leads to the finite jump in the Fermi momentum. In response the Hall coefficient $R_H\propto 1/x$ undergoes the corresponding  sudden jump.

As an illustration of the above consider the $T-B$ phase digram for YbRh$_2$Si$_{2}$ \cite{geg} (Fig. 1). Taking into account that the behavior of YbRh$_2$Si$_{2}$ strongly resembles the behavior of  YbRh$_2$(Si$_{0.95}$Ge$_{0.05}$)$_2$ \cite{geg1}, we can conclude that both the above described $T-B$ diagram and the behavior of the Hall coefficient $R_H$ are in good agreement with experimental facts \cite{geg,geg1}.

This work was supported by Russian Foundation for Basic Research, Grant No 05-02-16085.

\begin{figure}[!ht]
\begin{center}
\includegraphics[width=0.47\textwidth]{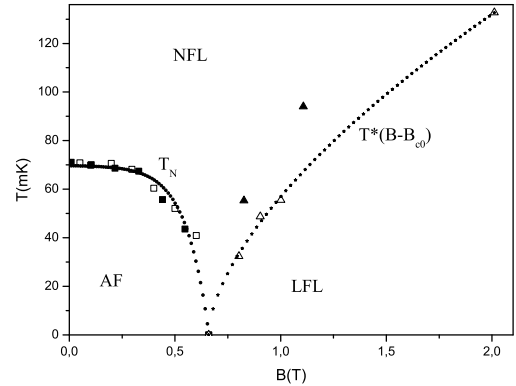}
\end{center}
\caption{Line separating AF and the NFL state is a guide to the eye, line separating the NFL and the LFL phases $T^*(B-B_{c0})\propto \sqrt{B-B_{c0}}$ \cite{shag4}.}
\label{fig1}
\end{figure}

\end{document}